\documentclass[12pt,preprint]{aastex}

\usepackage{graphicx}
\shorttitle{Galactic Plane Annihilation Radiation}
\shortauthors{Teegarden et al.}

\begin{document}

\title{INTEGRAL/SPI Limits  on Electron-Positron Annihilation 
Radiation from the Galactic Plane}

\author{B. J. Teegarden\altaffilmark{1}, K. Watanabe\altaffilmark{1}, P.
Jean\altaffilmark{2}, J. Kn\"{o}dlseder\altaffilmark{2},V. Lonjou\altaffilmark{2}, 
J. P. Roques\altaffilmark{2}, G. K. Skinner\altaffilmark{2},  P. von
Ballmoos\altaffilmark{2}, G. Weidenspointner\altaffilmark{2}, A.
Bazzano\altaffilmark{3}, Y. M. Butt\altaffilmark{4}, A.
Decourchelle\altaffilmark{5}, A. C. Fabian\altaffilmark{11}, A.
Goldwurm\altaffilmark{5}, M. G\"{u}del\altaffilmark{6}, D. C.
Hannikainen\altaffilmark{7}, D. H. Hartmann\altaffilmark{8}, A.
Hornstrup\altaffilmark{4}, W. H. G. Lewin\altaffilmark{9}, 
K. Makishima\altaffilmark{10}, A. Malzac\altaffilmark{11}, J. Miller\altaffilmark{4}, A. N.
Parmar\altaffilmark{12}, S. P Reynolds\altaffilmark{13}, R. E.
Rothschild\altaffilmark{14}, V Sch\"{o}nfelder\altaffilmark{15},   J. A. Tomsick\altaffilmark{14}, and J. Vink\altaffilmark{16}}

\altaffiltext{1}{Laboratory for High Energy Astrophysics, NASA/Goddard Space
Flight Center, Greenbelt, MD 20771, US}

\altaffiltext{2}{Centre d'Etudes Spatiale des Rayonnements, CNRS/UPS, BP 4346,
31028 Toulouse Cedex 4, France}

\altaffiltext{3}{Istituto di Astrofisica Spaziale e Fisica Cosmica
IASF, Roma, Italy}

\altaffiltext{4}{Harvard-Smithsonian Center for Astrophysics, 60
Garden Street, Cambridge, MA 02138}

\altaffiltext{5}{Service d'Astrophysique, Orme des Merisiers,
CE-Saclay, 91191 Gif-sur-Yvette, Cedex, France}

\altaffiltext{6}{Paul Scherrer Institut, WŸrenlingen \& Villigen, 5232
Villigen PSI, Switzerland}

\altaffiltext{7}{Observatory, PO Box 14, FIN-00014, University of
Helsinki, Finland}

\altaffiltext{8}{Clemson University, Clemson, SC 29634-0978, USA}

\altaffiltext{9}{Center for Space Research, Massachusetts Institute of
Technology, 77 Massachusetts Avenue, Cambridge, MA 02139-4307, USA}

\altaffiltext{10}{Department of Physics, University of Tokyo, 7-3-1 Hongo,
Bunkyo-ku, Tokyo 113-0033, Japan}

\altaffiltext{11}{Institute of Astronomy, Madingley Road, Cambridge,
CB3 0HA, UK}

\altaffiltext{12}{Astrophysics Division, Research and Scientific Support
Department of ESA, ESTEC, Postbus 299, 2200 AG Noordwijk, The
Netherlands}

\altaffiltext{13}{North Carolina State University, Department of Physics, Box
8202, Raleigh, NC 27695-8202}

\altaffiltext{14}{Center for Astrophysics and Space Science, University of
California at San Diego, La Jolla, CA 92093-0424}

\altaffiltext{15}{Max-Planck-Institut f\"{u}r extraterrestrische Physik, PO Box 1603, 85740 Garching, Germany}

\altaffiltext{16}{SRON Nat. Inst. for Space Research, Sorbonnelaan 2, 3584 CA Utrecht, The Netherlands}

\begin{abstract}
    The center of our Galaxy is a known strong source of
    electron-positron 511-keV annihilation radiation.  Thus far,
    however, there have been no reliable detections of annihilation
    radiation outside of the central radian of our Galaxy.  One of the
    primary objectives of the INTEGRAL (INTErnational Gamma-RAy Astrophysics Laboratory) mission, launched in Oct. 2002, is the detailed study of this radiation.  The Spectrometer
    on INTEGRAL (SPI) is a high resolution coded-aperture gamma-ray
    telescope with an unprecedented combination of sensitivity,
    angular resolution and energy resolution.  We report results from the first
    10 months of observation.  During this period a significant
    fraction of the observing time was spent in or near the Galactic
    Plane.  No positive annihilation flux was detected outside of the
    central region $ ( \mid l \mid \: > 40^{\circ } $) of our Galaxy.
    In this paper we describe the observations and data analysis
    methods and give limits on the 511-keV flux.
\end{abstract}


\section{Introduction}

Electron-positron annihilation radiation from the central region of
our galaxy was first reported by balloon-borne instruments more than
30 years ago \citep{joh72,joh73,hay75}.  Since the first detection
there have been many balloon and satellite observations
\citep{lev78,lev80,geh91,lev93,mah94,tee96,pur97}.  Both narrow-line
511-keV and continuum positronium emission have been observed
\citep{lev80,pur97,tee96,str03}.  Early measurements \citep{rie81,geh91,lev93} suggested
that the annihilation radiation was time-variable.  However, more
extensive later observations showed no evidence for variability
\citep{sha88,geh91,lev93,mah94,tee96,pur97,har98}.  The current consensus
favors the latter conclusion.  The most extensive results on the
spatial distribution of the annihilation radiation prior to INTEGRAL
came from the OSSE experiment on the Compton Gamma-Ray Observatory
(CGRO).  OSSE is a scintillator spectrometer with a simple $4^{\circ }
\times 11^{\circ } $ collimator.  Many scans were made through the
central region of our Galaxy, although relatively little data was
taken in the Galactic Plane outside of this region.  The OSSE data
were well described by a two-component bulge distribution plus a
somewhat less well established disk component.  One bulge model that
gave a reasonable fit to the data consisted of a circularly-symmetric
Gaussian of FWHM $ \sim 5^{\circ} $ plus an elongated Gaussian with a
width of $ \sim 30^{\circ} $ in longitude and $ \sim 5^{\circ} $ in
latitude \citep{kin01}.  Early OSSE papers discussed a possible
positive latitude enhancement, the so-called ``annihilation fountain''
\citep{pur97}.  However, later analysis \citep{mil01,kin01} found
little or no evidence for such a feature.

Many different ideas have been advanced, but despite numerous
observations over the past 30 years, the origin of the Galactic
positrons remains a mystery.  Plausible scenarios include (see \citet{cas04} and references therein):

\it Nucleosynthesis in Massive Stars: \rm Wolf-Rayet (WR) stars are massive stars
with strong stellar winds.  Radioactive elements created through hydrostatic nucleosynthesis in the cores of such stars can be convected to the surface and carried into the interstellar medium (ISM) by their stellar winds.  Some of the radioactive elements will undergo $ \beta^{+} $-decay, and the emitted positrons will annihilate in the interstellar medium  long before they can
escape from the Galaxy.

\it Type Ia Supernovae: \rm Radioactive nuclei are created by explosive
nucleosynthesis in Type Ia supernovae, some of which decay with
positron emission.  The positrons will annihilate either in the
expanding shell or the surrounding ISM. The principle uncertainty is
in the escape fraction of annihilation photons from the expanding
shell.

\it Hypernovae: \rm A hypernova is an asymmetric explosion of a WR star.  Positrons can be efficiently transported outward by strong jets and escape into low optical depth regions where they subsequently annihilate.

\it Black Holes and Pulsars: \rm Electron-positron pair production is
believed to occur in black hole jets and pulsar magnetospheres.  

\it Cosmic Ray Interactions: \rm Cosmic rays colliding with nuclei in the ISM can produce $\pi^+$\footnote{$\pi^+$ decays into $\mu^+$ which decays into  positron.} or excitation followed by $\beta^+$-decay.

\it Dark Matter: \rm There has been recent speculation on the possible
existence of light ($m <~100 \mbox{~MeV} $) dark matter particles
which would decay or annihilate primarily through the formation of electrons and positrons.  Dark matter decay or annihilation could possibly account for the bulge component of the
electron-positron annihilation radiation.

\section{Observations}

The Spectrometer on INTEGRAL (SPI) is a coded-aperture telescope using
an array of 19 cooled germanium detectors for high-resolution
spectroscopy (Vedrenne et al.  2003).  It covers the energy range $ 20
\-- 8000 \mbox{ keV} $ (with an energy resolution varying between $ 2
\mbox{ and } 6 $ keV), has an angular resolution of $ \sim 3^{\circ }
$ and a total field-of-view of $ \sim 25^{\circ } $ FWHM. One of its
primary objectives is the detailed study and mapping of diffuse
gamma-ray line emission.  The INTEGRAL observing program has two main
parts, the Core Program, which is essentially guaranteed time for the
teams of scientists who built and operate INTEGRAL, and the Open
Program which comprises competitively-selected observations of guest
investigators.  The Core Program includes a deep exposure of the
central radian of our Galaxy, repeated scans of the Galactic Plane and
selected point source observations.  A large fraction of the Open
Program observations were in or near (within $ \pm 20^{\circ } $) of
the Galactic Plane.  The coauthors of this paper include most of the
guest investigators selected for observations within $ \pm 20^{\circ }
$ of the Galactic Plane during the first year of INTEGRAL
observations.  We restrict ourselves in this paper to the Galactic
Plane outside of the central region ($ \mid l \mid \: > 40^{\circ }
$).  Results from the central region have been presented in other
papers \citep{jea03a,kno03,jea04,kno04,lon04,wei04a}.  Fig.  1 is an
exposure map of INTEGRAL observations during the period 2003 Dec to
2004 Oct.  The combined data set, including Core Program, Open
Program, and calibration observations amounts to 8.9 Msec exposure of
the Galactic Plane of which 2.1 Msec is in the region $ \mid l \mid
\: > 40^{\circ } $.  Solar and trapped particles can cause rapidly
changing instrument background levels that can be misinterpreted as
source flux.  The data were carefully screened to eliminate such
periods using a combination of on-board rates and data from the GOES
satellites.  Isolated points with large deviations that could be due
to either transmission or processing errors were also removed.

\section{Data Analysis}

In this paper we use the so-called ``light-bucket'' method of analysis
since we are interested mainly in broad-scale diffuse Galactic Plane
emission.  In this method all detectors are summed for each pointing,
which suppresses the imaging information on the $ 3^{\circ } $ scale
of the mask.  SPI is then a Òlight-bucketÓ collecting all photons in
its $ \sim 25^{\circ } $ FWHM field-of-view.  This mode is useful for
studying diffuse emission on spatial scales $ \ga 20^{\circ } $ and is
likely to be less sensitive to systematic effects than the full
imaging mode which requires an exact knowledge of the detailed
properties of the coded-aperture mask and the individual detectors.
We use the same Monte-Carlo-generated (GEANT) response function as is
used in the imaging mode \citep{stu03}, except that we sum the
response over all 19 detectors.  This work complements that of
\citet{kno04}, which concentrates on the imaging aspects of the
instrument, and provides an independent verification of some of the
conclusions reached there.

SPI has a high 511-keV background level due mainly to cosmic-ray
interactions in the instrument and surrounding spacecraft
\citep{jea03b,wei03,tee04}.  Some of these interactions make excited nuclei that undergo $\beta^+$-decay which produces a strong 511-keV background line.   The 511-keV signal-to-background ratio for observations in the central region of the Galaxy is typically only a few percent.  The INTEGRAL spacecraft is in a highly eccentric orbit with an initial apogee of 155000 km and perigee of 9000 km.  INTEGRAL
therefore spends most of its time outside of the magnetosphere,
which means that the background is relatively stable in comparison
with that for a low-earth orbiting instrument (e.g. OSSE, HEAO C-1).
The cosmic-ray flux typically varies by $ 5 \-- 10 \% $ over time
periods of the order of a year resulting in an instrumental background
at 511 keV whose variation is of the same order or larger than the
measured 511-keV flux.  A precise determination of this background is
therefore critical to extracting the optimum performance from SPI. In
this analysis we have modeled the 511-keV background using various
on-board tracers of the primary cosmic-ray intensity
\citep{jea03b,lon04,tee04}.  Two of these, the rate of saturated events in
the germanium detectors (GEDSAT) and the plastic-anticoincidence rate
(PSAC), have been found to be the most useful.  The former samples
cosmic-rays $ \ga 200 $ MeV and the latter $ \ga 10 $ MeV. We assume
that far from the Galactic Plane ($ \mid b \mid \: > 20^{\circ } $)
there is no significant 511-keV emission and use this data (3.1 Msec live time) from the first year of operation to determine the best-fit background model.
If there were quasi-isotropic 511-keV emission it would be suppressed
in our analysis.  We calculated the total 511-keV line counting rate  by integrating over a 10-keV interval centered at 511.0 keV and subtracted a continuum determined by a simple interpolation between two 10-keV windows on either side of the line.   We fit this data and found that a linear combination of the GEDSAT and PSAC rates is a good background predictor, but that there is a significant long-term monotonically increasing residual in the fit.

A detailed Monte Carlo model for SPI has been implemented under the
GEANT software package.  An enhanced version of this package MGGPOD
\citep{wei04b} was used that treats gamma-ray background production in
a much more extensive and complete manner.  The Monte Carlo
background simulations are not accurate enough to make absolute
predictions of the background levels.  However, they are useful in
identifying the dominant isotopes for background production and their
half-lives.  Long half-life decays can account for the sort of
residual described in the preceding paragraph.  The Monte Carlo
analysis identified $ ^{65} $Zn, a spallation product of germanium with a
half-life of 244 days, as a possibly significant contributor to the
511-keV background.  We included a term in the model for this decay
channel and found that it significantly improved the fit although the actual value of the half-life was not well-constrained by the available data.  $^{22}$Na with a 2.6 yr half-life is another possible contributor to the long-term 511-keV build-up, however its intensity in the simulation
is low. The accuracy of the model 511-keV background prediction is estimated to be
0.3 \%.

\section{Results}

Fig.  2 is a plot of the background-subtracted SPI 511-keV counting rate in the $ -10^{\circ
} < b < 10^{\circ } $ band as a function of Galactic Longitude.    A clear peak centered at $ l = 0^{\circ } $ is visible.  The first papers of SPI results \citep{jea03a,kno03} reported that this enhanced flux region was well represented by a circularly-symmetric Gaussian
with a width of $ 7^{\circ } \-- 10^{\circ } $.  More recent work
\citep{kno04} has found evidence for an asymmetry in the
bulge with a latitude width of $ 13^{\circ } \pm 5^{\circ } $ and
longitude width of $ 25^{\circ } \pm 4^{\circ } $.  This spatial
distribution appears to be qualitatively similar to the two-component
bulge plus disk description of the OSSE results \citep{kin01}.  Our
Òlight-bucketÓ analysis is not very sensitive to these differences in
morphology.  As a check on our method we have fit the data with a
circularly-symmetric Gaussian centered at $ l = 0^{\circ } $, $b =
0^{\circ } $ and a fixed width of $ 10^{\circ } $.  We derive a total
flux of $ \sim 1 \times 10^{-3} \mbox{ cm}^{-2} \mbox{ s}^{-1} $,
which is consistent with early SPI results (Jean et al.  2003a, 2004).  Other authors
\citep{sha88,kin01} have reported higher fluxes from the central region.  A possible explanation for these differences is the existence of an extended emission halo to which INTEGRAL might not be sensitive.  Resolution of this question must await the accumulation of more data and better coverage of regions of the sky away from the Galactic Center.

Outside of the $ \mid l \mid \: < 40^{\circ } $ central region there
is no evidence for any significant 511-keV emission.  The error bars
vary widely due to the non-uniform coverage of the Galactic Plane.
They are smallest in regions where the exposure is deepest, near the
Crab ($l = 185^{\circ} $), Cygnus ($ l = 71^{\circ} $), Vela ($ l =
-94^{\circ } $) and Cas A ($ l = 116^{\circ} $).  In Fig. 3 the
511-keV counting rate is plotted as a function of Galactic latitude
for two longitude regions, $ -180^{\circ } < l < -40^{\circ} $ and $
40^{\circ} < l < 180^{\circ} $.  Again there is no evidence for any
significant flux.  Even if there were errors in our background model,
we would not expect them to be correlated with the SPI pointing
direction.  The absence of any \it relative \rm enhancement in the
latitude distributions in Fig.  3 is strong evidence for the absence
of detectable 511-keV emission from the Galactic Plane.

Upper limits (90 \% confidence) for the 511-keV flux in the Galactic
Plane are given in Table~1.  The plane outside of the central region
has been divided into $ 10^{\circ} \times 10^{\circ} $ tiles.  Within
each tile the SPI response to a point source anywhere in the tile has
been calculated.  The average value for this response was used to
determine the upper limit values of Table~1.  We estimate a systematic
uncertainty in the flux of $ \sim 7 \times 10^{-5} \mbox{ cm}^{-2}
\mbox{ s}^{-1} $ (due to the uncertainty of the background model) which is added in quadrature to the statistical error
in calculating the upper limits.  There is significant variation in
the values due to the varying exposure.  The lowest values ($ \sim 1
\times 10^{-4} \mbox{ cm}^{-2} \mbox{ s}^{-1} $) are of the order of
10 \% of the total flux in the central bulge component.  The lowest
upper limits are at locations of high exposure where well-known
sources in the Galactic Plane were observed for calibration and/or
scientific study.  

We have used a second method to search for large-scale 511-keV
emission from the Galactic Plane .  We have performed global fits to
the entire $ ( \mid l \mid \: >  40^{\circ })$ data set using several
source distributions.  The following emissivity distributions were
used:

\it Old Disk \rm \citep{rob03}:

\begin{equation}
\rho(R,z) = \rho_{0}(e^{-(a/R_{0})^{2}} - e^{-(a/R_{i})^{2}})
\end{equation}

where $ a = R^{2} + z^{2}/\epsilon^{2} $, $\epsilon = 0.14 $,
$ R_{0} = 5 \mbox{ kpc (disk scale radius)} $,

$ R_{i} = 3 \mbox{ kpc (inner disk truncation radius)} $.

\it Young Disk \rm \citep{rob03}:

\begin{equation}
\rho(R,z) = \rho_{0}(e^{-(0.25 + (a/R_{0})^{2})^{1/2}} - e^{-(0.25 + 
(a/R_{i})^{2})^{1/2}})
\end{equation}

where $ \epsilon = 0.0791, R_{0} = 2.53 \mbox{ kpc}, R_{i} = 1.32 \mbox{ 
kpc} $.

These emissivities were integrated along the line-of-sight to produce
spatial flux distributions.  For the sake of completeness we also included 
a flat distribution (in galactic longitude with a Gaussian latitude 
profile of FWHM  $ 5^{\circ} $).  The fits to these distributions all
produce results that are consistent with zero 511-keV flux from the
Galactic Plane (outside of the central region).  The 90~\%-confidence upper limits are: old disk, $4.4 \times 10^{-4} \mbox{ cm}^{-2} \mbox{ s}^{-1} $; young disk, $ 3.9
\times 10^{-4} \mbox{ cm}^{-2} \mbox{ s}^{-1} $; flat disk, $ 5.0
\times 10^{-4} \mbox{ cm}^{-2} \mbox{ s}^{-1} $.

\section{Conclusions}

We have analyzed data from the first 10 months of the INTEGRAL mission
using the so-called ``light bucket'' method, which in principle is
relatively less sensitive to systematic effects than the more standard
imaging techniques.  We have found no significant 511-keV flux from
the Galactic Plane ($ -20^{\circ } < b < 20^{\circ } $) outside of the
central region ($ -40^{\circ } < l < 40^{\circ } $).  Our method of
analysis is sensitive to both point and diffuse sources.  The values
of the limits in $ 10^{\circ } $ tiles vary between $ 1.2 \times
10^{-4} \mbox{ cm}^{-2} \mbox{ s}^{-1} \mbox{ and } 1.9 \times
10^{-3} \mbox{ cm}^{-2} \mbox{ s}^{-1} $ due to large variations in
the exposure in different regions of the Galactic Plane.  We have also
performed global fits of the entire data set which set limits on the
total 511-keV flux in the Galactic Plane.  These values limit the
broad-scale 511-keV emission in the outer region of the Plane to $< 40
- 50 \% $ of that from the central region ($  \mid l \mid \: <
40^{\circ } $).  Over the course of the mission (expected to last $ >
5 $ yr.)  the exposure will deepen and likely become more uniform,
which will lead to better and more uniform limits and, perhaps, to
detections.

\acknowledgements

INTEGRAL is a project of the European Space Agency to which NASA is a contributing partner.


\clearpage                                                                                                                                                                                     
\begin{figure}
\plotone{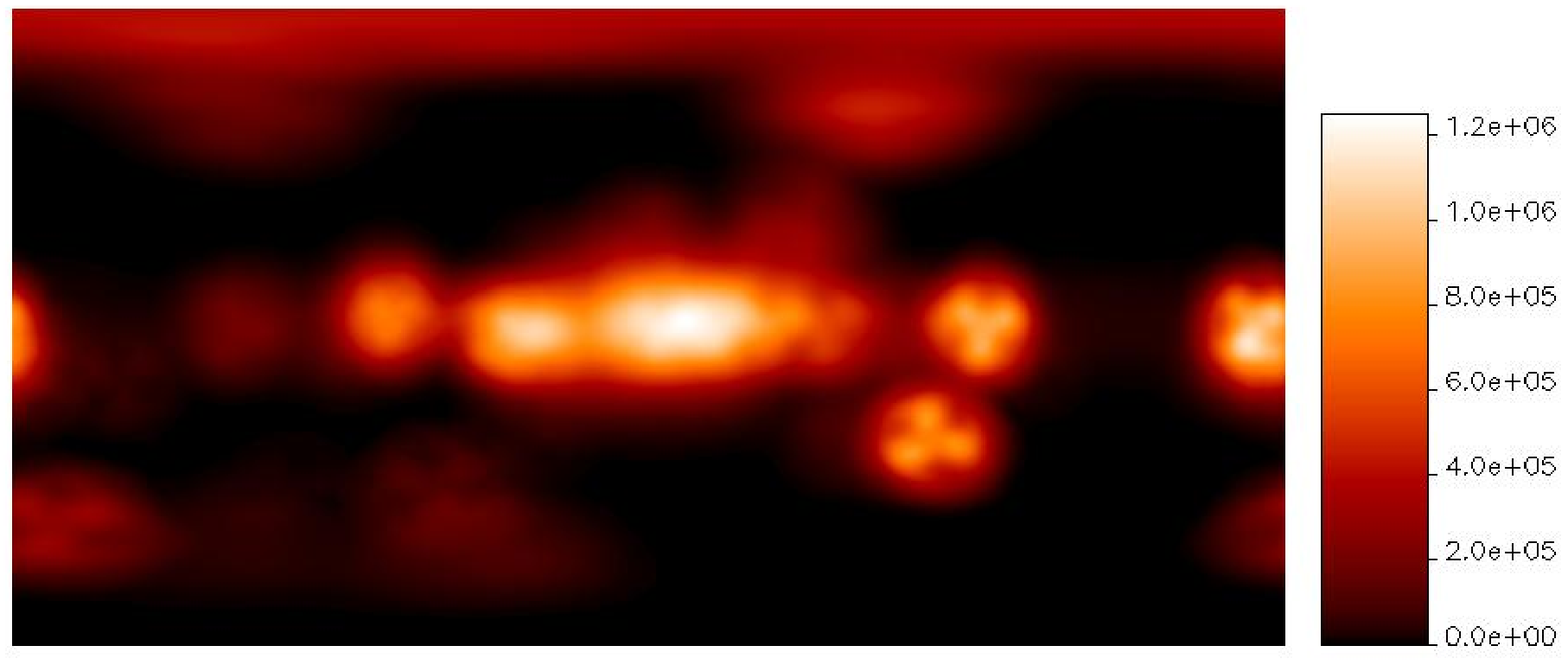}
\epsscale{.5}
\caption{INTEGRAL/SPI Exposure map of the Galaxy
for the first 10 months of operation of INTEGRAL. Horizontal range is
$ 180^{\circ} \mbox{ to } -180^{\circ} $ .  Vertical range is $
-90^{\circ} \mbox{ to } 90^{\circ} $. A large fraction of the
observations are concentrated in the Galactic Plane.  The
characteristic ÒpinwheelÓ patterns in some regions are due to
modulation of the light bucket response by the SPI coded-aperture
mask.  Color bar units are seconds.}
\end{figure}

\clearpage
\begin{figure}
\plotone{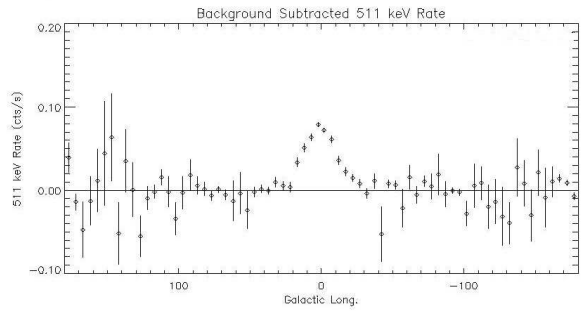} 
\caption{INTEGRAL/SPI background-subtracted 511-keV
counting rate as a function of Galactic Longitude for the region $
-20^{\circ } < b < 20^{\circ } $.  The significant excess centered at
$ l = 0^{\circ } $ is mainly due to the 511-keV bulge component.  No
significant flux is seen in the region $  \mid l \mid \: > 40^{\circ
} $. The large differences in error bars are due to widely varying
exposures in different parts of the Galactic Plane.}
\end{figure}

\clearpage
\begin{figure}
\epsscale{1}
\plottwo{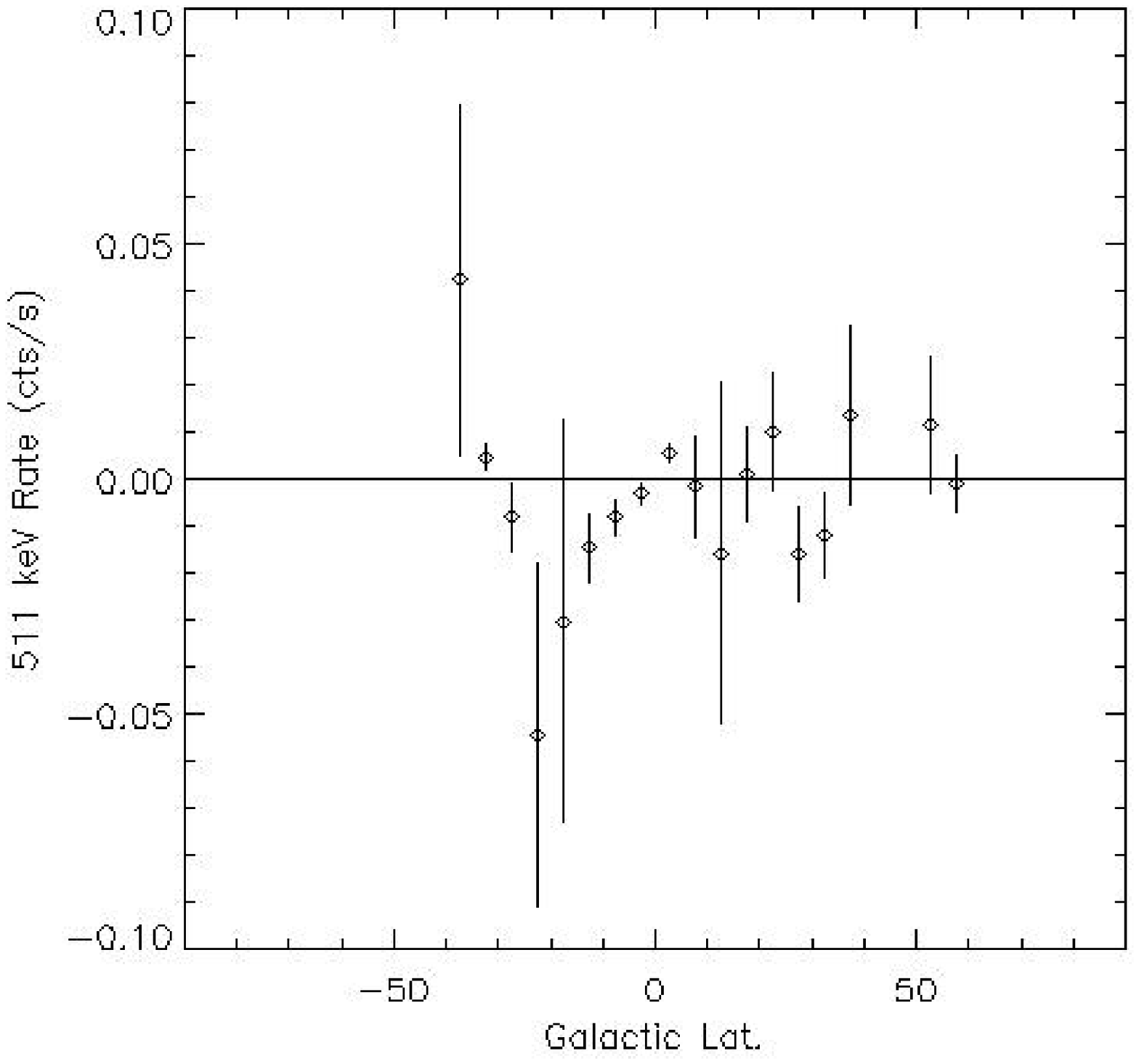}{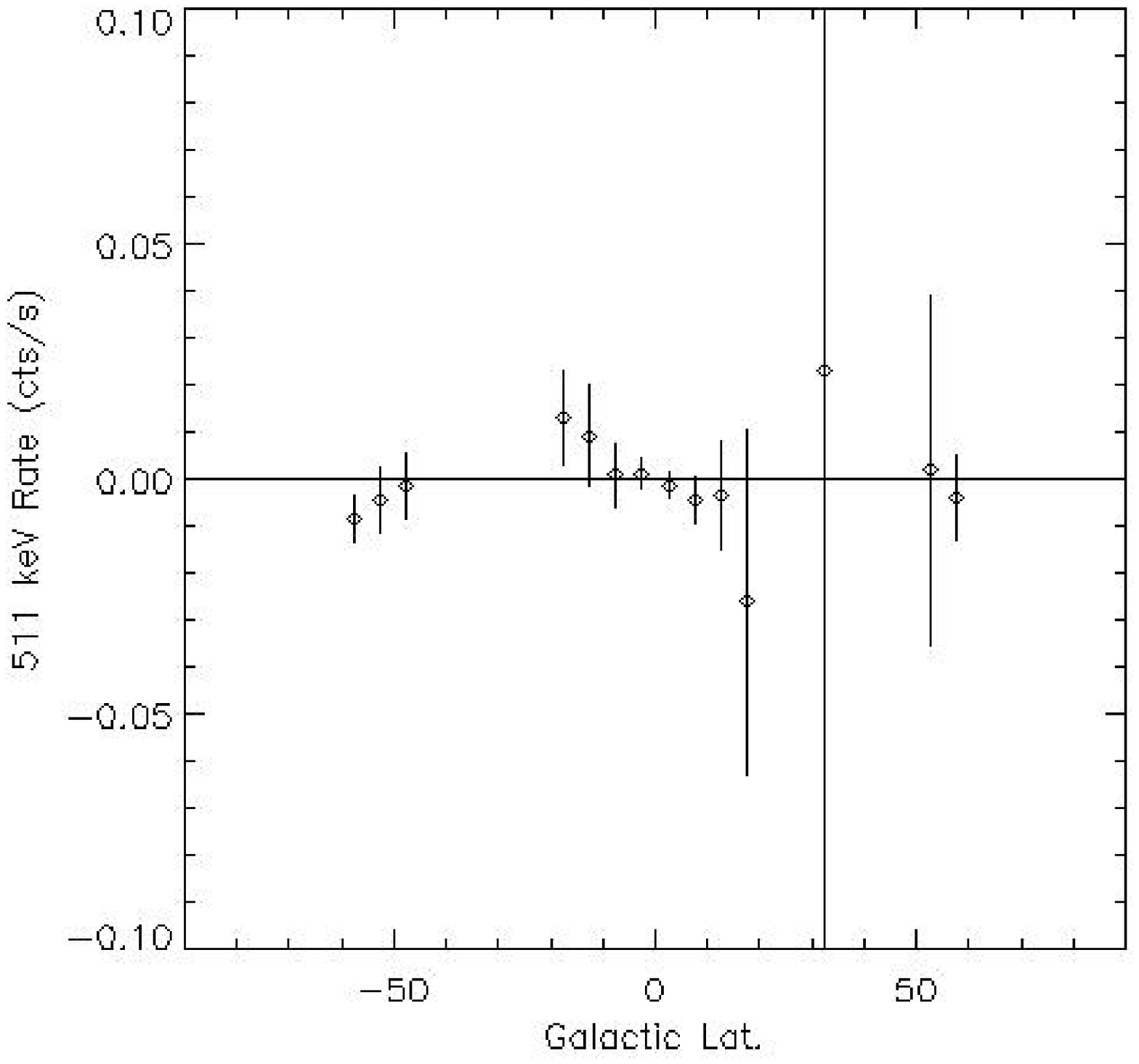} 
\caption{Left panel, INTEGRAL/SPI
background-subtracted 511-keV counting rate as a function of Galactic
Latitude for the region $ -180^{\circ } < l < -40^{\circ } $.  Right
panel, same as left but for the region $ 40^{\circ } < l < 180^{\circ
} $.  No significant flux excess in the plane is seen.}
\end{figure}


\clearpage
\begin{deluxetable}{rrrrrrrrrrrrrrr}
\tabletypesize{\scriptsize}
\tablenum{1a}
\tablecaption{Galactic Plane 511-keV Flux Limits}
\tablewidth{0pt}

\tablehead{ \colhead{l/b} & \colhead{-175} & \colhead{-165} &
\colhead{-155} & \colhead{-145} & \colhead{-135} & \colhead{-125} &
\colhead{-115} & \colhead{-105} & \colhead{-95} & \colhead{-85} &
\colhead{-75} & \colhead{-65} & \colhead{-55} & \colhead{-45} }

\startdata 
-15 & 1.31 & 1.51 & 2.89 & 8.05 & 8.55 & 9.49 & 4.45 & 2.38 & 1.82 &
1.81 & 2.07 & 2.12 & 2.21 & 2.58 \\
 -5 & 1.17 & 1.36 & 2.53 & 3.42 & 3.71 & 3.42 & 2.35 & 1.41 & 1.12 &
1.29 & 1.98 & 2.51 & 2.14 & 1.91 \\
  5 & 1.62 & 2.10 & 4.03 & 3.54 & 3.46 & 3.17 & 2.66 & 1.32 & 1.13 &
1.24 & 2.53 & 2.79 & 2.12 & 2.00 \\
 15 & 5.11 & 5.79 & 11.63 & 9.02 & 7.23 & 6.73 & 6.30 & 3.47 & 1.82 &
3.75 & 8.17 & 5.37 & 2.74 & 2.17
\enddata

\end{deluxetable}

\begin{deluxetable}{rrrrrrrrrrrrrrr}
\tabletypesize{\scriptsize} 
\tablenum{1b} 
\tablecaption{Galactic Plane 511-keV Flux Limits (cont.)}
\tablewidth{0pt}

\tablehead{ \colhead{l/b} & \colhead{45} & \colhead{55} &
\colhead{65} & \colhead{75} & \colhead{85} & \colhead{95} &
\colhead{105} & \colhead{115} & \colhead{125} & \colhead{135} &
\colhead{145} & \colhead{155} & \colhead{165} & \colhead{175} }

\startdata 
-15 & 1.97 & 5.13 & 5.98 & 4.68 & 4.45 & 4.64 & 5.39 & 4.33 & 5.51 &
6.46 & 4.64 & 4.11 & 2.86 & 1.48 \\
 -5 & 1.20 & 1.52 & 1.44 & 1.26 & 1.75 & 2.02 & 2.32 & 1.89 & 1.89 &
3.00 & 4.32 & 3.46 & 2.07 & 1.33 \\
  5 & 1.18 & 1.37 & 1.21 & 1.14 & 1.39 & 2.07 & 2.37 & 2.05 & 1.94 &
3.54 & 5.31 & 4.44 & 3.28 & 1.92 \\
15 & 2.34 & 3.22 & 1.64 & 1.38 & 1.95 & 4.31 & 5.43 & 4.74 & 4.68 &
10.44 & 18.55 & 11.35 & 6.20 & 6.59 \enddata 

\tablecomments{Upper limits are 90 \% confidence values in $ 10^{\circ
} \times 10^{\circ } $ tiles centered on the given l,b values.  Flux
units are $ \mbox{ cm}^{-2} \mbox{ s}^{-1} \times 10^{-4} $}.

\end{deluxetable}

\end{document}